\documentclass[epj]{svjour}
\usepackage{graphicx}
\begin{document}
\title{ Bounded confidence model on a still growing scale-free network}
\author{A.O. Sousa \thanks{email:sousa@ica1.uni-stuttgart.de}}
\institute{Institute for Computational Physics (ICP), University of
  Stuttgart, Pfaffenwaldring 27, 70569 Stuttgart, Germany}
\date{Received: date / Revised version: date}
%
\abstract{
A Bounded Confidence (BC) model of socio-physics, in which the agents have 
continuous opinions and can influence each other only if the distance 
between their opinions is below a threshold, is simulated on a still growing 
scale-free network considering several different strategies: for each new 
node (or vertex), that is added  to the network all individuals of the 
network have their opinions updated following a BC model recipe. The 
results obtained are compared with the original model, with numerical 
simulations on different graph structures and also when it is
considered on the usual fixed BA network. In particular, the
comparison with the latter leads us to conclude that it does not matter
much whether the network is still growing or is fixed during the
opinion dynamics.
} 
\maketitle

\section{Introduction}
It has recently been found that many systems, ranging from social science to 
biology, from economics to technology, which can be described as complex 
networks, seem to share some important topological features such as the 
scale-free degree distribution, where the probability that a node of these 
networks has $k$ connections follows a power-law $P(k) \sim k^{-\gamma}$, 
with $\gamma$ laying in a quite wide interval $2\leq\gamma\leq3$ 
\cite{vasquez}. For example, the Internet \cite{inter}, in which nodes 
are computers and routers and edges are physical or wireless connections 
between them; the World Wide Web \cite{www}, in which the nodes are the 
web pages (documents) and the edges are the hyperlinks that point from 
one document to another; the scientific 
collaboration network \cite{scient}, in which nodes are scientists and edges 
represent collaboration in scientific papers (two scientists are linked if 
and only if they are co-authors of the same paper). 

Of particular interest here are the social networks, where the nodes 
are people, and the ties between them are (variously) acquaintance, 
friendship, political alliance or professional collaboration. More 
specifically, in this paper we study a model of opinion dynamics 
evoluted on a scale-free network. Many models about opinion dynamics 
have been proposed
\cite{axelrod,deffuant,krause,bc,latane,sznajd,pseudo,bonne,santo,dittmer}, 
however, in general only binary opinions were
considered \cite{bc,latane,sznajd,pseudo,bonne}, such way we have 
only minority 
or majority opinions and it is impossible to distinguish between 
moderate and extreme opinion. As simplistic as it appears, the binary 
decision framework has been used to address surprisingly complex 
problems \cite{latane}. An interesting and straightforward extension 
would be to consider continuous opinions, i.e., a wide spectrum of 
opinions \cite{deffuant,krause,santo,stauffer}. Modeling of such a model 
was earlier started by applied mathematicians and focused on the 
conditions under which a panel of experts would reach a consensus 
\cite{krause,bc,mat}.

In the present work, we simulate on the Barab\'asi-Albert network a simple 
consensus-finding model, $\,$ the ``Bounded Confidence'' (BC) model 
\cite{deffuant}, where originally the individuals (sites or nodes) 
living in the continuum, in contrast to our network, have continuous 
opinions and the individuals can influence each other only if the distance 
between their opinions is below a threshold. 

In the next section, we describe the standard Bounded Confidence model and 
the Barab\'asi-Albert network, in section 3, we present our results and in 
section 4, our conclusions.

\begin{figure*}[!htb]
\begin{center}
\includegraphics[angle=0,scale=1.0]{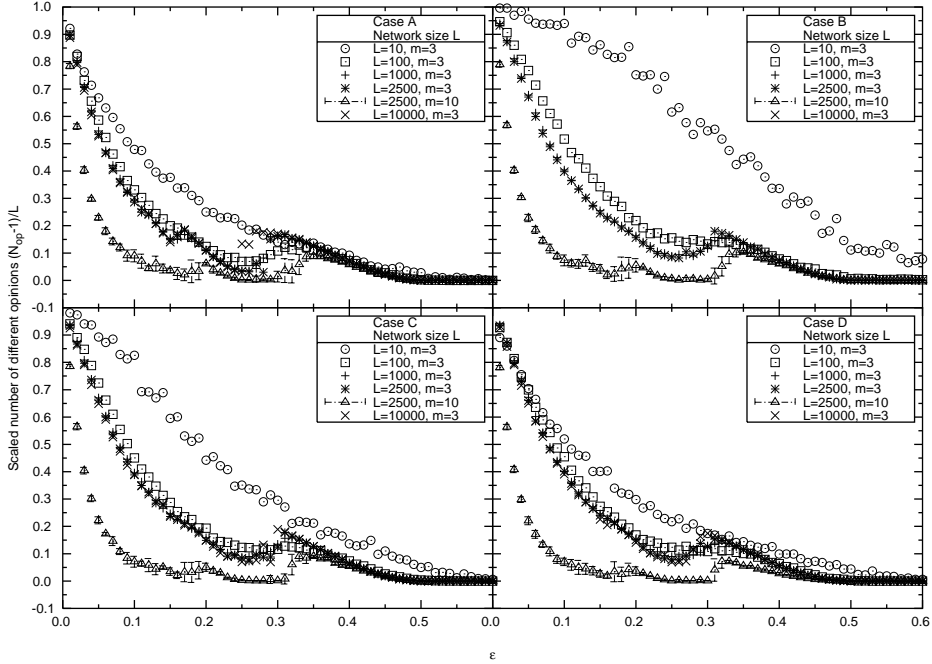}
\end{center}
\caption{Scaled total number of different opinions versus the inverse of the 
constant confidence bound $\epsilon$ for the {\bf Case A} (top left), 
{\bf Case B} (top right), {\bf Case C} (bottom left) and {\bf Case D} (bottom
right) using different network sizes.}
\end{figure*}

\section{The Model}
The Barab\'asi-Albert (BA) network \cite{BA} starts to grow from an initial 
cluster of $m$ fully connected sites. Each new node (or vertex), that is 
added to the network creates $m$ links (edges, ties) that connect it to 
previously added nodes. The power-law distribution emerges as a result of 
preferential attachment, which means that the probability of a new link to 
end up in a vertex $i$ is proportional to the connectivity $k_i$ of 
this vertex. The validity of the preferential attachment was confirmed 
within real networks analysis \cite{BA}. The BA algorithm generates 
networks with the desirable scale-free distribution $P(k) \propto
k^{-3}$ and small values of the average shortest path. The only
striking discrepancy between the BA model and real networks is that
the value of the clustering coefficient - which is the probability
that two nearest neighbors of the same node are also mutual 
neighbors - predicted by the theoretical model decays very fast 
with network size and for large systems is typically several orders 
of magnitude lower than found empirically.

In the Bounded Confidence model \cite{deffuant,krause,dittmer} 
each individual (or node) is represented by a continuous opinion $s_i$, whose 
initial value is a random number chosen between zero and one ($0<s_i<1$). At 
every time step two randomly chosen individuals $i$ and $j$ have their 
opinions readjusted only if their difference in opinion, 
$\delta_{ij} = (s_i - s_j)$, is smaller in magnitude than a threshold 
$\epsilon: |\delta_{ij}| < \epsilon$. In this way, the opinions are 
adjusted according to:
\begin{equation}
s_i = s_i - \mu \delta_{ij} \hspace{1cm} {\rm and} \hspace{1cm}
s_j = s_j + \mu \delta_{ij}
\label{eq:change}
\end{equation} 
\noindent where $\mu$ is the convergence parameter whose values may range 
from 0 to 0.5 and characterizes the flexibility in changing the opinion. 
It has been observed that the qualitative dynamics mostly depend on the 
threshold $\epsilon$, which controls the number of peaks in the final 
distribution of opinions. The number $L$ of individuals and the
parameter $\mu$ only influence convergence time and the width of the 
distribution of final opinions \cite{deffuant}. 

At each time step $0 < t \le L-m$, we have the following process:
\begin{enumerate}
\item The BA network grows, i.e., one new site $i$ (individual) is
  added, and a random opinion $s_i$ ($ 0 < s_i < 1$) is set as initial 
opinion to the single new node $i$ of the network.

\item For each new site added to the network, $N_d=10^6$ BC runs are 
performed. For each run, all the nodes are randomly visited and updated 
(a random list of nodes assures that each node is reached exactly once) 
by selecting a node $i$ at random and, among its connected nodes, a 
node $j$ at random. If $|\delta_{ij}|<\epsilon$, their opinions, $s_i$ 
and $s_j$, are  re-adjusted  (Eq. \ref{eq:change}).
\end{enumerate}
In contrast to a recent work simulating the BC model on the
usual fixed BA network \cite{hmo,weis}, the consensus process of the BC 
model is not performed after the complete network had been
constructed, but while the network grows, i.e, while each new node is 
added to the network, a Bounded Confidence prescription is applied: 
the already existing sites have their opinions readjusted every time 
when a new site is added. Notice that this assumption has already been 
used before, however in the context of a binary opinion model on a 
Barab\'asi-Albert (BA) network \cite{bonne} and on a deterministic 
pseudo-fractal network \cite{pseudo}.

The following four cases have been investigated:
\begin{itemize}
  \item {\bf Case A:} For each selected site $i$, {\bf all} the nodes 
    connected to it are randomly visited and tested.

  \item {\bf Case B:} For each selected site $i$, {\bf only one} neighbor 
    $j$ is selected from the $m$ sites which $i$ had selected to make 
    a link when it was added to the network. If $|\delta_{ij}|>\epsilon$, 
    another $i$ is picked up. 
    
  \item {\bf Case C:} For each selected site $i$, {\bf only one} neighbor 
    $j$ is taken from {\bf all} its $k_i$ neighbors. If 
    $|\delta_{ij}|>\epsilon$, another $i$ is picked up.

  \item {\bf Case D:} For each selected site $i$, {\bf only one} neighbor 
    $j$ is taken from {\bf all} its $k_i$ neighbors. In case of 
    $|\delta_{ij}|>\epsilon$, another neighbor $j$ is randomly picked up. 
    If after $k_i$ times no neighbor $j$ provides $|\delta_{ij}|<\epsilon$, 
    then another $i$ is selected.
 \end{itemize}

\begin{figure*}[!ht]
\begin{center}
\includegraphics[angle=0,scale=1.0]{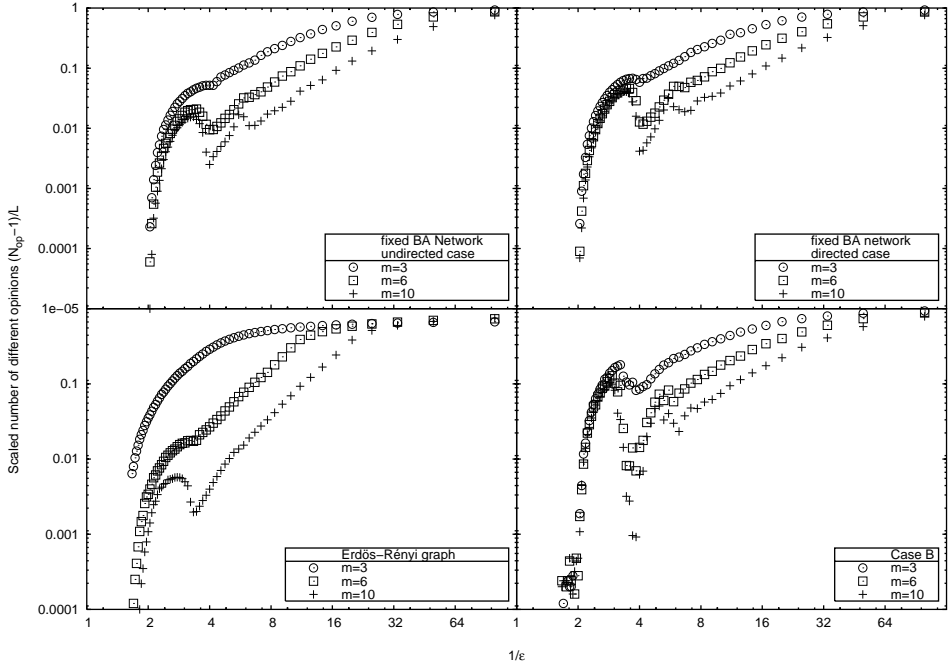}
\end{center}
\caption{Scaled total number of different opinions versus the inverse of the 
constant confidence bound $\epsilon$ for the {\bf undirected case} (top left), 
{\bf directed case} (top right), {\bf Erd\"os-R\'enyi} (bottom left) and 
{\bf Case B} (bottom right) using different values of $m$. Both axes are 
logarithmic.}
\end{figure*}

\section{Results}
After $t$ time-steps the network has $L=m+t$ nodes (individuals). The curves 
presented here correspond to the results averaged over $100$ samples. The 
opinions are placed in bins of width $10^{-6}$ and are counted by checking 
which bins are occupied and do not have the lower neighboring bin occupied. 
In this way, the total number of fixed opinions is obtained.

In Figure 1: case A (top left), case B (top right), case C (bottom left) 
and case D (bottom right), we present the total number of different fixed 
opinions $N_{op}-1$ divided by the network size $L$ as a function of 
the inverse of the constant confidence bound $\epsilon$. It has been 
observed in all cases (Fig. 1) that when $\epsilon > 0.5$ a full 
consensus (only one opinion survives) is reached and for $\epsilon < 0.5$, 
no consensus is reached and the number of different fixed opinions 
increases with decreasing $\epsilon$. In fact, when $\epsilon$ goes 
to zero, the number of surviving opinions approaches the network size $L$, 
as expected \cite{hmo}. The threshold value of $\epsilon$ obtained in our 
simulations ($\epsilon > 0.5$, a complete consensus and $\epsilon < 0.5$, no
consensus) is in a complete concordance with the original BC model
\cite{deffuant,naim} and with recent numerical simulations of it on 
different graph structures \cite{santo} that provide strong numerical evidence 
that this value does not depend on the way the agents are connected 
to each other (i.e, the graph structure), but it relays on the social 
dynamics. The small statistical error of the threshold $\epsilon$ for 
complete consensus in our simulations (shown only for $L=2500$ and $m=10$ 
in Fig.1) confirms the good agreement of our results with the previous ones 
\cite{deffuant,naim,santo}.

We can also observe that when the network size $L$ increases, the total
number of different fixed opinions $N_{op}-1$ increases $\propto L$,
while the scaled number of different opinions - $(N_{op}-1)/L$ -
for smaller system sizes $L$ ($L \le 100$) presents stronger finite 
size effects, however, they become weaker for larger system sizes $L$, 
then the $L$-dependence of the scaled excess number almost disappear.

\begin{figure}[b]
\begin{center}
\includegraphics[angle=0,scale=0.7]{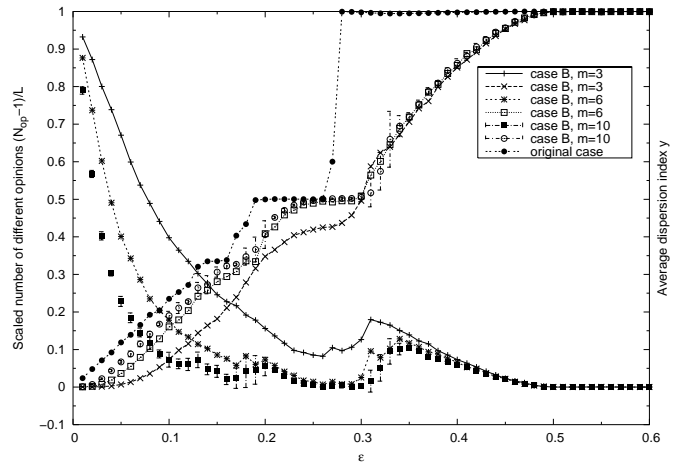}
\end{center}
\caption{Scaled total number of different opinions and the average dispersion 
index y as a function of the constant confidence bound $\epsilon$. $L=2500$ 
(Case B) and $L=10000$ (Original model). Errorbars are shown for $m=10$.}
\end{figure}

Moreover, for large $L$, interestingly we have observed in all cases that the 
number of final opinions reaches a local minimum around $\epsilon = 0.25$ 
(a steeper one) and $\epsilon \approx 0.15$ (a smoother one). Besides, as 
the connectivity "m" increases the minimum becomes deeper and others minimums 
appear (see all cases in Fig.1 when $L=2500$ and $m=10$). These local
minima could not be noticed before in the usual fixed BA network 
studied in Ref.\cite{hmo}, due to the lack of detailed simulations for 
different values of the threshold $\epsilon$. In this way, in order 
to analyze more carefully
these minima we have also performed simulations of the BC model on the usual 
fixed BA network, on the Erd\"os-R\'enyi random graph, as well as without 
any network topology, i.e., the original BC model. In the former one, as 
it was previously studied in \cite{hmo}, two different cases have been 
investigated: the directed case (our case B) and the undirected one 
(our case C).
  
The construction of the Erd\"os-R\'enyi random graph \cite{random} starts 
with a set of $L$ isolated vertices, then successive edges are randomly 
added with a probability $p$. In this way, the total number of edges is 
$m_{t}=pL(L-1)/2$ and the average number of neighbors of a node 
(degree or connectivity) is $m=p(L-1)$. In the limit $L \rightarrow \infty$, 
the mean number of bonds per site can be approximated by $pL$ and a 
Poissonian connectivity distribution is observed. In our simulations, the 
graph has been built in such way that each node has at least $m$
links. Moreover, on this topology we have investigated only the case C.

Figure 2 shows the scaled total number of different opinions versus the 
inverse of the threshold $\epsilon$ for the usual fixed BA network 
\cite{hmo}: undirected case (top left) and directed case (top right), 
for the Erd\"os-R\'enyi (bottom left) and for case B on a still 
growing BA network (bottom right) 
when different values of $m$ are considered. As we can see, the large is the 
average connectivity, which is made by increasing $m$, the steeper are the
local minima observed when $\epsilon \approx 0.25$ and 
$\epsilon \approx 0.15$ (it does not seem to appear for the  Erd\"os-R\'enyi
graph). An individual (node) with few connections (link) has less chances to
interact with a neighbor whose opinion is close enough ($s_i - s_j<\epsilon$) 
to its own opinion to actually interact, in this way, many of the individuals
remain outside the distribution of clustered opinions. This implies that if 
the average connectivity increases more individuals should converge into 
the same opinion's cluster, which make us to expect that in the limit of large
$m$ (everybody interacts with everybody) the results become closer to those 
of the original BC model (without any network topology): the majority of the
individuals has the same opinion or only one cluster is observed for 
$\epsilon > 0.25$ \cite{weis}.

In the original BC model, it has been found that for large $L$ the number 
of clusters varies as the integer part of $1/2\epsilon$ \cite{deffuant}. 
In such way that for $\epsilon>0.25$ most of individuals belong to 
only one single cluster, and for $\epsilon<0.25$ several large ones 
(see original case in Fig. 3) \cite{weis}. On the other hand, however
simulations on the BA network show the results follow this $1/2\epsilon$ 
rule, the existence of many individuals with lower connectivity in scale-free
networks makes the fraction of individuals into the same cluster to become 
smaller \cite{weis}. Nevertheless, it is important to emphasize that since 
the opinion are represented by real numbers, the convergence towards to 
a full consensus (only one opinion) inside a cluster is never actually 
reached, i.e., the clusters correspond to a group of individuals with 
very similar but not exactly equal opinions. The differences between the 
clusters are related to the threshold $\epsilon$.

A simple way to check clustering and specially its average, is the dispersion
index $y$ proposed by Derrida and Flyvberg \cite{derrida}:
\begin{equation}
y=\frac{\sum_i {s_i}^2 }{(\sum_i s_i)^2}
\label{eq:index}
\end{equation}
\noindent where $s_i$ is the cluster size, i.e., the number of individuals 
in each cluster $i$ and $L$ the total number of individuals. In Figure 3, 
we plot the scaled final number of opinions 
and the dispersion index $y$ against the threshold $\epsilon$ for different 
values of the connectivity $m$. When $0.10<\epsilon<0.15$, 
$0.15<\epsilon<0.2\,$ and $0.25<\epsilon<0.3$ local minima can be observed, 
and $\epsilon>0.5$ a full consensus is reached. As we can notice, the regions 
corresponding to these minima are the same ones where distinct behaviors 
in the dispersion index $y$ are observed, which seem to be related to the 
transition of the number of clusters: four to three large clusters, three 
to two, two to one, respectively. In the original case (filled circles in 
Fig. 3) the transition regions are more sharper than in those ones 
observed for the case B, in which the dispersion index $y$ varies more 
smoother and is indicated by a slope becoming the steeper the larger the
connectivity $m$ is. These results are in a good agreement with previous 
ones for the original BC model and for the BC model on the usual BA network 
\cite{weis}. The index $y$ varies smoothly as a function of the threshold 
$\epsilon$ due mainly to the existence of many individuals with lower
connectivity $m$ that remain outside of the opinion convergence process 
and do not cluster. In this way, in an infinite network and in the limit of 
large mean connectivity $m$, one would get a sharp step function in these
transition regions for the mean dispersion index $y$ versus the threshold 
$\epsilon$, i.e., the results for the scale-free network becomes very similar
to those ones obtained for the original BC model, where everybody interacts 
with everybody \cite{weis}. Moreover, the continuous increase of the index 
$y$ as a function of the threshold $\epsilon$ when $\epsilon>0.3$, while 
for the original case $y$ is equal to unity (corresponding a full consensus 
of the system, i.e, all the individuals belong to the same opinion cluster), 
shows clearly the existence of many individuals kept out of the clustering 
process \cite{weis} in the BA scale-free network.

\section{Conclusions}
Using a consensus model with bounded confidence on a still growing 
Barab\'asi-Albert network, we have shown that the system reaches a
full consensus when $\epsilon > 0.5$ and for $\epsilon < 0.5$, no 
consensus is reached anymore and the number of different fixed opinions
increases with decreasing $\epsilon$. This critical value for finding
a complete consensus is the same one obtained in the original
random case when the individuals were considered to live in the 
continuum \cite{deffuant,naim,weis}, as well as when the BC model was 
simulated on the usual fixed BA network \cite{weis} and for numerical 
simulations of the BC model on several graph structures \cite{santo}. 
Once the BC prescription here is performed 
while the network grows, so when a new node (individual) is added, it can 
find the existing sites of the network already in a complete consensus. Thus, 
we believe that this particular feature seems to be responsible for reducing 
strongly the finite size effects related to the $L$-dependence observed in 
\cite{hmo}. We have also found local minima in the scaled final number 
of opinions as a function of the threshold $\epsilon$, which are related 
to the phase transition in the number of opinion clusters. To investigate 
this question further, it has been performed additionally simulations of 
the original BC model, as well as considering it on the usual BA network 
(directed and undirected case) and on the Erd\"os-R\'enyi random graph 
for several values of the confidence bound parameter $\epsilon$. All 
studied cases independently of the underlying topology show these local 
minima, which occur in the phase transition regions of the number of the 
opinions clusters as a function of the threshold $\epsilon$. In particular 
for the small-world graphs, one observes that the minima become steeper 
and a sharper step function for the dispersion index $y$, the larger mean 
connectivity is, and also, in the limit of large connectivity and large 
network size one would get the same results obtained for the original 
BC model. Moreover, the reason to have the highest values of the dispersion 
index $y$ smaller than unity (as obtained in the original BC model) is 
due to the fact that only a fraction of all individuals belongs to the 
big opinion cluster(s) resulting from the convergence process 
\cite{deffuant,weis}. In summary, our results lead us to conclude that 
it does not matter much whether the network is still growing or is 
fixed during the opinion dynamics, the opinion spreading 
properties remain the same. An identical conclusion has also been found 
for computer simulations on binary opinion dynamics \cite{pseudo,bonne}.

\begin{acknowledgement}
The author thanks D. Stauffer and R.V. Mesquita for helpful discussions and 
critical reading of the manuscript, and the important suggestions of the
anonymous referees.
\end{acknowledgement}


\begin{thebibliography} {99}
\bibitem{vasquez} A. V\'asquez, M. Boguna, Y. Moreno, R. Satorras, A. 
Vespignani, Phys. Rev. E {\bf 67}, 046111 (2003)
 
\bibitem{inter} M. Faloutsos, P. Faloutsos and C. Faloutsos, Comp. Comm. 
Rev. {\bf 29}, 251 (1999); R. Albert, H. Jeong and A. L. Barab\'asi, Nature
{\bf 401}, 130 (1999). 

\bibitem{www} R. Albert, H. Jeong and A.L. Barab\'asi, Nature {\bf 401},
  130 (1999); R. Kumar, P. Raghavan, S. Rajalopagan and A. Tomkins,
  Proceedings of the 9th ACM Symposium on Principles of Database Systems, p.1 
(1999).
 
\bibitem{elegans} J.G. White, E. Southgate, J.N. Thompson and S. Brenner, 
Phil. Trans. R. Soc. London {\bf 314}, 1 (1986); D.J. Watts and S.H. Strogatz,
Nature {\bf 393}, 440 (1998); L.A.N. Amaral, A. Scala, M. Barthelemy and 
H.E. Stanley, Proc. Natl. Acad. Sci. USA {\bf 97}, 11149 (2000).

\bibitem{scient} M.E.J. Newman,  Proc. Natl. Acad. Sci. USA {\bf 98},
  404 (2001); M.E.J. Newman,  Phys. Rev. E. {\bf 64}, 016131 (2001); H. Jeong, 
Z. N\'eda and  A.L. Barab\'asi,  cond-mat/0104131 (2001); F. P\"utsch, Adv. 
Compl. Sys. {\bf 6}, 477 (2003). 

\bibitem{axelrod} R. Axelrod, J. Conflict Resolut. {\bf 41}, 203 (1997).  

\bibitem{deffuant} G. Deffuant, D. Neau, F. Amblard and G. Weisbuch, Adv. 
Complex Syst. {\bf 3}, 87 (2000); G. Deffuant, F. Amblard, G. Weisbuch 
and T. Faure, JASSS {\bf 5}, issue 4, paper 1 (2002) 
(http://jasss.soc.surrey.ac.uk/5/4/1.html).

\bibitem{krause} R. Hegselmann and U. Krause, JASSS {\bf 5}, issue 3, 
paper 2 (2002) (http://jasss.soc.surrey.ac.uk/5/3/2.html).

\bibitem{naim} E. Ben-Naim, P.Krapivsky, S. Redner, Physica D {\bf 183}, 190 (2003).

\bibitem{santo} S. Fortunato, Int. J. Mod. Phys. C. {\bf 15}, issue 9,
  in press (2004) = cond-mat/0406054.

\bibitem{stauffer} D. Stauffer, A.O Sousa and C. Schulze, JASSS 
{\bf 7}, issue 3, paper 7 (2004) (http://jasss.soc.surrey.ac.uk/7/3/7.html) = 
cond-mat/0310243.

\bibitem{bc} S. Chatterjee and S. Seneta, J. Appl. Prob. {\bf 14}, 89 (1977);
J. Cohen, J. Kajnal and C.M. Newman, Stochastic Process and their Applications 
{\bf 22}, 315 (1986); N.E. Friedkin and E.C. Johnsen, Adv. in Group Processes 
{\bf 16}, 1 (1999);.

\bibitem{latane} B. Latan\'e and A. Nowak, {\bf 13}, 43 in {\it Progress in 
Communication Sciences}, edited by G.A. Barnett and F.J. Boster, Ablex 
Publishing Corporation (1997); K. Kacpersky and J. Holyst, Physica A 
{\bf 287}, 631 (2000); S. Galam, J. Stat. Phys. {\bf 61}, 943 (1990); S. 
Galam, Physica A {\bf 238}, Physica A {\bf 238}, 66 (1997); S. Galam
and S. Wonczak, Eur. Phs. Journal B {\bf 18} 183 (2000).

\bibitem{sznajd} K. Sznajd-Weron and J. Sznajd, Int. J. Mod. Phys. C
  {\bf 11}, 1157 (2000); D. Stauffer, A.O. Sousa and S. Moss de
  Oliveira, Int. J. Mod. Phys. C  {\bf 11}, 1239 (2000); F. Slanina 
and H. Lavi\'cka, Eur. Phys. J. B, 35, 279 (2003)= cond-mat/0305102; 
D. Stauffer, JASSS {\bf 5}, issue 1, paper 4 (2002)
(http://jasss.soc.surrey.ac.uk/5/1/4.html).

\bibitem{pseudo}M.C. Gonzalez, A.O. Sousa and H. Herrmann,
  Int. J. Mod. Phys. C {\bf 15}, 45 (2004) = cond-mat/0307537.

\bibitem{bonne} J. Bonnekoh, Int. J. Mod. Phys. C. {\bf 14}, 
1231 (2003) =  cond-mat/0305125.

\bibitem{dittmer} J.C. Dittmer, Nonlinear Analysis {\bf 47}, 4615 (2001).

\bibitem{mat} M. Stone, Ann. Math. Stat. {\bf 32}, 1369 (1961); 
J.F. Laslier, Econ. Appl. {\bf 42-3}, 155 (1989).

\bibitem{BA} A.L. Barab\'asi and R. Albert, Science  {\bf 286}, 509
  (1999); H. Jeong, B. Tombor, R. Albert, Z.N. Oltvai and A.L. Barab\'asi, 
Nature {\bf 407}, 651 (2000); R. Albert and A.L. Barab\'asi, Rev. Mod. Phys. 
{\bf 74}, 47 (2002).

\bibitem{hmo} D. Stauffer and H. Meyer-Ortmanns, cond-mat/0308231 = 
Int. J. Mod. Phys. C. {\bf 15}, 241 (2004).

\bibitem{random} P. Erd\"os and A. R\'enyi, Publicationes Mathematicae 
{\bf 6}, 290 (1959). 

\bibitem{weis} G. Weisbuch, Eur. Phys. J. B {\bf 38}, 339 (2004); G. Weisbuch, 
G. Deffuant and F. Amblard, cond-mat/0305125 (2004);

\bibitem{derrida} B. Derrida, H. Flyvberg, J. Phys. A {\bf 19}, L1003 (1986).

\end{thebibliography}
\end{document}